\documentclass{projectXtemplate}
\usepackage[utf8]{inputenc}
\usepackage[english]{babel}
\usepackage{graphicx}
\graphicspath{{./images/}}
\usepackage[bookmarks=false]{hyperref}
\usepackage[backend=biber, sorting=none]{biblatex}
\usepackage{upquote}
\usepackage{amsmath}
\usepackage{amssymb}
\usepackage{mathtools}
\usepackage{comment}
\usepackage{bm}
\usepackage{float}
\usepackage[caption = false]{subfig}
\pdfoutput=1

\bibliography{biblio}
  {\begin{list}{}%
          {\setlength{\leftmargin}{#1}}%
          \item[]%
  }
  {\end{list}}

\usepackage{calrsfs}
\DeclareMathAlphabet{\pazocal}{OMS}{zplm}{m}{n}

\title{Topological Data Analysis (TDA) Techniques \\
Enhance Hand Pose Classification from \\ 
ECoG Neural Recordings}

\author{\textbf{Simone Azeglio$^{1, \dag}$, Arianna Di Bernardo$^{1, \dag, *,}$, Gabriele Penna$^2$,  Fabrizio Pittatore $^2$, Simone Poetto$^1$,}}
\affil{\textbf{Johannes Gruenwald$^3$, Christoph Kapeller$^3$, Kyousuke Kamada$^4$, and Christoph Guger$^{3,5}$}}

\affil{$^1$Master Student, Department of Physics, University of Turin, Italy}
\affil{$^2$Master Student, Biomedical Engineering, Polytechnic University of Turin, Italy}
\affil{$^3$g.tec medical engineering GmbH, Schiedlberg, Austria}
\affil{$^4$Neurosurgery, Megumino Hospital, Eniwa, Japan}
\affil{$^5$Guger Technologies OG, Graz, Austria}
\affil{*E-mail: arianna.dibernard@edu.unito.it}

\thisvolume{XX}
\thismonth{October}
\thisyear{20XX}
\thisdoi{XX}
\thisstartpage{01}
\thisendpage{03}

\begin{document}

\maketitle
\thispagestyle{thefirstpage}

\begin{center}
   ABSTRACT 
\end{center} 

\vspace{-2mm}
\blfootnote{\dag These authors contributed equally}
\textbf{Electrocorticogram (ECoG) well characterizes hand movement intentions and gestures. In the present work we aim to investigate the possibility to enhance hand pose classification, in a Rock-Paper-Scissor - and Rest - task, by introducing topological descriptors of time series data. We hypothesized that an innovative approach based on topological data analysis can extract hidden information that are not detectable with standard Brain Computer Interface (BCI) techniques. To investigate this hypothesis, we integrate topological features together with power band features and feed them to several standard classifiers, e.g. Random Forest, Gradient Boosting. Model selection is thus completed after a meticulous phase of bayesian hyperparameter optimization. With our method, we observed robust results in terms of accuracy for a four-labels classification problem, with limited available data. Through feature importance investigation, we conclude that topological descriptors are able to extract useful discriminative information and provide novel insights. Since our data are restricted to single-patient recordings, generalization might be limited. Nevertheless, our method can be extended and applied to a wide range of neurophysiological recordings and it might be an intriguing point of departure for future studies.}

\section{\textbf{INTRODUCTION}}


\subsection{\textbf{ECoG}}

A Brain Computer Interface (BCI) is a device that records the electrical activity of the brain and outputs it directly to the computer in order to be collected and analysed. Nowadays, a wide range of BCI techniques are employed for different purposes, ranging from medical applications to education and gaming \cite{6165246}.
There are many types of BCI, working on different spatial and temporal resolution. BCI based on electroencephalogram (EEG) are the most widely used. This is due to the low cow cost and non-invasive set up as well as the high temporal resolution. They are based on the extraction of features from the EEG signal by processing event-related potentials (ERP) or oscillatory activity such as event-related synchronization/desynchronization (ERS/ERD) in the low frequency bands, up to 50 Hz. The most common type of feature based on oscillatory activities is the power band feature, which represents the power of a certain frequency range for each channel \cite{Lotte_2018}.
However EEG based BCI have their weakness in the low spatial resolution and low spatial to noise-ratio \cite{kapeller2014single}. Furthermore, the electrical potentials measured in the EEG suffer from a “smearing” phenomenon caused by the different electrical conductivities of the layered structure of the head \cite{hari2017meg}.
Invasive techniques such as electrocorticography (ECoG) or micro-electrode arrays (MEAs) \cite{Bout_2016} can overcome these issues. In particular ECoG provides a better signal quality \cite{li2017gesture} and yields a higher spatial resolution than EEG thanks to the small exposure diameter and higher density of the electrodes, as well as its higher invasiveness. Furthermore ECoG signal contains information up to 500 Hz, allowing the study of the power-bands in the high gamma activation (HGA) range, over 40 Hz \cite{kapeller2014single}.

\subsection{\textbf{A Multiclass Classification Problem}}
A key problem in BCI-based neural prosthesis control is decoding movement intentions from brain electrical activity. ECoG signals contain rich information correlated with motor activities, in particular with regards to hand gesture decoding \cite{PanGang2018RDoH}. Such a problem has received a lot of attention recently \cite{gruenwald2019time},\cite{Chestek_2013}, and can be seen in form of a multiclass classification problem. In this context, each hand gesture corresponds to a class, and the recorded ECoG signals are the objects to be classified. Previous studies approached this issue as a 3 class classification problem, classifying only the movement trials and ignoring the rest state \cite{gruenwald2019time}. In our work we'll deal with 4 classes, considering the rest state as an indipendent class. \\

The standard technique used to face such a classification problem is a combination of common spatial patterns (CSP) and multi-class linear discriminant analysis (LDA) \cite{kapeller2014single}, or multi-class support vector machine (SVM) \cite{LiYue2017GDUE},\cite{Taku2011electro}. Other studies suggest to use recurrent neural networks to exploit the temporal information in ECoG signals \cite{PanGang2018RDoH}.\\


In this study, we propose a Topological Data Analysis (TDA) based approach, that in combination with standard techniques from the BCI field, aims to provide a robust hand gesture decoding.

\section{\textbf{MATERIAL AND METHODS}}

\subsection{\textbf{Subject}}

For this study, we evaluated ECoG recordings of one patient (male, 26 years old), who suffered from intractable epilepsy and thus underwent surgical treatment at Asahikawa Medical University, Asahikawa, Japan. In the course of this treatment, a total of 140 ECoG electrodes were implanted on his right hemisphere. From these 140 electrodes, a high-density grid of 60 electrodes (Unique Medical Co., Ltd.; diameter 1.5 mm, spacing 5 mm, geometry 6 $\times$ 10) covered sensorimotor areas. We used these 60 electrodes for further processing. 

The patient voluntarily participated in research experiments besides the standard clinical procedure. These research experiments were approved by the institutional review board of Asahikawa Medical University and received certificate number 245 in June 2012. The patient provided written informed consent before participating in the experiments. For additional details (such as detailed electrode placement), we refer to \cite{gruenwald2019time}, where this patient appears as S2.

\subsection{\textbf{Experimental protocol}}

We here analyze ECoG data acquired during a hand motor experiment inspired by the famous hand game \emph{rock-paper-scissors}. Specifically, we instructed the patient to form one of the three hand poses (rock, paper, or scissors) according to a visual stimulus presented on a computer screen. These visual stimuli were presented to the patient for one second each and were interleaved by a distorted image shown for a randomized duration between 1.5 and 2.5 seconds. While the distorted image was shown, the patient was asked to return into a relaxed hand position. Consequently, this experiment constitutes a three-class motor control BCI experiment. We analyzed 30 trials per class for this subject in total.

\subsection{\textbf{Data Acquisition}}

We used a \emph{g.HIamp} biosignal amplifier (g.tec medical engineering GmbH, Austria) to digitize the neural recordings at a sampling rate of 1.2 kHz. For stimulus presentation, we employed the \emph{g.HIsys Highspeed Online Processing} toolbox (g.tec medical engineering GmbH, Austria) in the \emph{Simulink} environment (The MathWorks, Inc., Massachusetts, USA). We saved the raw data as \emph{MATLAB} files on a hard drive for further processing.

\subsection{\textbf{Preprocessing}}
The first operation performed was the re-referencing of the data by applying a common average reference (CAR) spatial filter, presented in \cite{art1}. The CAR filter subtracts, for every time-point and every channel, the mean of all the non-excluded channel, as in \cite{gruenwald2019time}. 
\newline The powerline frequency (50 Hz) was eliminated using a cascade of Notch Butterworth filters of order 5 up to the sixth harmonic, as in \cite{gruenwald2019time}.No channels selection strategy were applied. 
\newline After literature studies \cite{jiang2020power} we evaluated the signal in different bands. The ECoG signal presents information content up to 500 Hz \cite{staba2002quantitative}. For the feature extraction discussed in point F based on the Topological Data Analysis (TDA) method, different frequency bands were evaluated: (i) 1-500 Hz signal in order to evaluate also the low-frequency band (1-50 Hz), removing the DC offset, (ii) 100-500 Hz signal to evaluate only the high-gamma frequency information content, (iii) 50-300 Hz signal to assess a different bandcut which allows the evaluation of the high gamma frequency range band, as done in \cite{gruenwald2019time}, (iv) 1-500 Hz signal not filtered with the spatial CAR filter.\\


A further preprocessing step was completed for preparing the signals for the TDA phase, but not for the power band extraction one. When dealing with computational TDA techniques, signals of equal length are required. \textit{Zero-padding} was employed to uniform variable length signals: zero values was appended to the shorter vectors. The padding was applied to the end of the vectors.

\subsection{\textbf{Feature Extraction: Power Band}}

For the power band features extraction only the CAR filter and the Notch filter were applied. To extract the power band features three different bandwidth were used: 60-90 Hz, 110-140 Hz and 160-190 Hz as in \cite{kapeller2014single}. The bandpass windows were chosen in order to avoid the powerline interference. A 4th order Butterworth filter was used for each temporal filter $T_{fb}$.
\begin{equation}
y(t)=T_{fb}[x(t)]
\end{equation}
The filtered signals were then segmented into 2s epochs, each corresponding to a rest/movement trial. To estimate the band power each channel was squared and averaged over a 2s window. The features were then logarithmically scaled \cite{sig_process}.
\begin{equation}
f_{pb}=\log \left(\langle[y(t)]_{n}^{n+fs\cdot 2}\rangle^{2}\right)
\end{equation}
where $\langle  \cdot  \rangle$ represents the mean value.
A total of 180 features were extracted, 1 per each channel for each frequency band used.


\subsection{\textbf{Feature extraction: Topological Data Analysis}}

\textit{Topological Data Analysis} (TDA) refers to a collection of methods used to extract geometric features from complex data \cite{WassermanLarry2018TDA}. By leveraging algebraic topology and computational geometry it is possible to discover structures in data, that are relevant and robust to noise.\\

A popular method in TDA is \textit{Persistent Homology} (PH), a technique that analyzes the shape of the data to deduce their intrinsic properties, like holes or connected components 
\cite{Edelsbrunner_persistenthomology}.\\
PH can be applied to an heterogeneous variety of datasets (images, graphs, time series) by approximating them into elementary objects that preserve their topological properties, and referred to as \textit{simplicial complexes} \cite{edelsbrunner2010computational}.\\

In this section we aim to introduce the basic tools utilized in PH. A detailed description of this research area can be found in \cite{BiasottiS2008Dsbg}, \cite{CarlssonGunnar2014Tprf}.

\paragraph{Point clouds.}

In the present work, the intent is to apply PH to preprocessed and segmented ECoG signals. In order to use topological tools on such a multivariate time series, we converted them into point clouds, a suitable data type for PH \cite{WuChengyuan2021Tmlf}. \\

To do this we use a modified version of the Takens delay embedding, taking inspiration from \cite{WuChengyuan2021Tmlf}. 
This method consists of sampling the time series at fixed time steps, de facto constructing a point cloud. The Takens embedding is indeed defined on univariate time series, and it depends on three parameters: 
\begin{itemize}
    \itemsep0em
    \item the \textit{time delay} $\tau$ is the time between two consecutive values for constructing one embedded point; 
    \item the \textit{dimension} $N$ represents the dimension of the embedding space;
    \item the \textit{stride} $s$ indicates the duration between two consecutive embedded points.
\end{itemize}

Here we consider the multivariate time series \mbox{ $\mathbf{X}=\{  \mathbf{x}(t_i) \}_{i=1,..,T}$ }, where $\mathbf{x}(t_i) = (x^1(t_i), x^2(t_i),.. x^d(t_i)) \in \mathbb{R}^d$, $d$ is the number of 1D time series and $T$ is the lenght of the series. The stride parameter $s$ determines the sampled times: $s=t_{i+1}-t_{i}$.
By applying the Takens' embedding prevoiusly introduced, for a sampled value $t_n$, we obtain the corresponding point cloud $\mathbf{X}_n$, defined as follow:

\begin{equation}\label{eq1}
\mathbf{X}_n=\left(\mathbf{x}({t_{n}}), \mathbf{x}({t_{n}+\tau}), \ldots, \mathbf{x}({t_{n}+(N-1) \tau})\right)
\end{equation}

consisting of $N$ points in $\mathbb{R}^{d}$.\\

In the case under consideration, having multivariate time series consisting of $60$ channels, we set the dimension parameters to $1$. In this way, from every channel, one coordinate value is picked, resulting in a total of $60$ coordinates for the embedded points.\\

The number of points in every point cloud is determined by the stride parameter and by the length of each samples. In the present work we set $s=10$ and $\tau=1$. We then obtained a single point cloud for each ECoG signal; where each point cloud is intended as a collection of $240$ $60$-dimensional points.
 


\paragraph{Simplices and Simplicial Complexes.}

Once the point clouds were computed for each ECoG sample, we extracted their topological features via PH. The standard technique requires the conversion of the point clouds into \textit{simplicial complexes}, combinatorial structures that preserve the topological properties of the clouds. To be defined, we need the notion of \textit{k-simplex} $\sigma$, a convex hull of $k+1$ affinely independent points $ \{  v_{0}, v_{1}, \ldots, v_{k}  \} \in \mathbb{R}^k $:
\begin{equation}
  \sigma=\left[v_{0}, v_{1}, \ldots, v_{k}\right].  
\end{equation}

A face $\mu$ of a \textit{k-simplex} is a simplex of lower dimension, and we write $\mu \subseteq \sigma$. For instance, a 0-simplex is a point, a 1-simplex a segment, a 2-simplex a triangle, and both the 0-simplex and the 1-simplex are faces of the 2-simplex.\\

A simplicial complex $K$ is then a finite set of simplices such that for every $\mu \subseteq \sigma \in K $, we have $\mu \in K$, and for every $\sigma_1, \sigma_2 \in K$, $\sigma_1 \cap \sigma_2 $ is either empty or a face of both.\\

Finally, we recall the concept of \textit{filtration} of a simplicial complex $K$, defined as a nested sequence of complexes $\emptyset=K^{0} \subseteq K^{1} \subseteq \cdots \subseteq K^{m}=K$. Then we refer to $K$ as a \textit{filtered complex}. \\


When dealing with point cloud data $\mathbf{X}_n$, the most intuitive type of simplicial complex is the  \textit{Vietoris-Rips complex}, i.e. the simplicial complex whose $k$-simplices are determined by each subset of $k+1$ euclidan point which are pairwise within distance $\varepsilon$:

\begin{equation}
V R_{\varepsilon}(\mathbf{X}_n)=\{S \subseteq \mathbf{X}_n / d(x(t_n), x(t_m)) \leq 2 \varepsilon, \forall x(t_n), x(t_m) \in S\}.
\end{equation}

In the present work, we computed the Vietoris-Rips complexes for each point cloud previously defined in (\ref{eq1}), and then we applied PH.

\paragraph{Homology.}

Homology is a general technique for associating a sequence of algebraic objects (usually abelian groups) to a topological space $K$. The so obtained sequence of homology groups $H_k(K)$ provides information about the number of k-dimensional "holes" in $K$ for every dimension $k=0,1,2...$.

To define the notion of Homology we first introduce the concepts of \textit{chain group} and \textit{boundary operator}.\\

A \textit{k-chain} is a formal sum of $k$-simplices in $K$:

\begin{equation}
c=\sum_{i=1}^{k} a_{i} \sigma_{i}, \ a_{i} \in \{ 0,1 \}.
\end{equation}

The set of the $k$-chain in $K$ together with the addition operation form the free abelian group $C_k(K)$, referred to as \textit{k}-th chain group.\\


The boundary operator $\partial_{k}: C_{k}(K) \rightarrow C_{k-1}(K)$ is defined on an oriented simplex $\sigma=\left[v_{0}, v_{1}, \ldots, v_{k}\right]$ by

\begin{equation}
\partial_{k}(\sigma)=\sum_{i=0}^{k}(-1)^{i}\left[v_{0}, \ldots, \hat{v}_{i}, \ldots, v_{k}\right] ,
\end{equation}
where the notation $\hat{v}_{i}$ indicates that the element $v_i$ is excluded from the sum.\\

In addition we define the \textit{cycle group} $Z_k(K)$ and the \textit{boundary group} $B_k(K)$, both subgroups of $C_k(K)$: 

\begin{equation}
\begin{aligned}
Z_{k}(K) &=\operatorname{ker} \partial_{k} \\
B_{k}(K) &=\operatorname{Im} \partial_{\mathrm{k}+1}
\end{aligned}
\end{equation}

Finally, we obtain the \textit{k-th Homology group} as the quotient group:
\begin{equation}
H_{k}(K)=Z_{k}(K) / B_{k}(K).
\end{equation}
The \textit{Betti number} $\beta_k$ is the rank of the $k$-th homology group: 

\begin{equation}
\beta_{k}=\operatorname{rank}\left(H_k (K)\right)
\end{equation}

and indicates the number of $k$-dimensional holes in the simplicial complex $K$. For instance, $\beta_0$ counts the number of connected components, $\beta_1$ the number of circular holes and $\beta_2$ the number of cavities in $K$.

\paragraph{Persistent Homology.}

The concept of Persistent Homology makes the notion of Homology suitable for computational approaches. The advantage of this technique is the possibility to investigate the homology groups of a topological space at multiple levels of scale.\newline
Given a filtered complex $K$, PH attempts to identify the topological features of $K$ that \textit{persist} along the filtration.\\
For each inclusion $K^{i} \hookrightarrow K^{j}$ in the filtration, a  homomorphism is naturally induced for each dimension $k$:
\begin{equation}
f_{k}^{i, j}: H_{k}\left(K^{i}\right) \rightarrow H_{k}\left(K^{j}\right)
\end{equation}

Then, \textit{k-th persistent homology group} is defined as:
\begin{equation}
H_{k}^{i, j}(K)= \operatorname{Im} f_{k}^{i, j},
\end{equation}
and the \textit{persistent Betti number} is its rank:
\begin{equation}
\beta_{k}^{i, j}=\operatorname{rank}\left(H_{k}^{i, j}(K)\right).
\end{equation}

The $k$-th persistent homology of a filtered simplicial complex gives more refined information than just the homology of the single subcomplexes: structures on data come on multiple scales and can be nested or in more complicated relationships \cite{Edelsbrunner_persistenthomology}.\\

We can visualize the information given by the $k$-th persistent homology group by drawing the following \textit{Persistence diagram}.\\ The interval $[i, j)$ indicates the lifetime of a $k$-homology class across the filtration: the endpoints of the interval represent the steps of the filtration at which the $k$-homology class born ($i$) or died ($j$), while its difference $l^{i,j}_k=j-i$ represents the \textit{persistence} of the $k$-homology class.\\
Such an interval can be represented as the point $(i, j)$ in the Euclidean plane $\mathbb{R}^2$. All points live in the half-space above the diagonal, and the persistence is easily visible as the vertical distance to the diagonal.

\paragraph{Topological Features.}
Once the persistence classes of the point clouds $\mathbf{X}_n$ and the persistence diagrams were computed for each homology dimension $k$, we extracted the \textit{topological features}. In the present work, we focused on three classes of features, defined below.\\ 

The most intuitive feature we computed was the \textit{Number of Points} in the persistence diagrams, one for each homology dimension that we considered.\\

We then computed the \textit{Amplitude of a persistence diagram}, defined as its distance to the empty diagram, containing only the diagonal, and it is computed according to the chosen metric (\textit{Wasserstein}, \textit{Bottleneck}, \textit{Betti},  \textit{Landscape}, \textit{Heat}).\\

Finally, the \textit{Persistence entropy} measures the entropy of the points in a given persistence diagram. It is computed by taking the Shannon entropy of all persistences in the persistence diagram \cite{garin2019topological}:

\begin{equation}
E_k=\sum_{[i,j)} \frac{l^{i,j}_{k}}{L_k} \log \left(\frac{l^{i,j}_{k}}{L_k}\right)
\end{equation}

where $L_k$ is the sum of all persistences in the diagram: $L_k=\sum_{[i,j)}l^{i,j}_{k}$.\\

In \textit{Table \ref{table:ID_to_Feature}} Topological Features are mapped into an integer identifier (ID), which has been used in the implementation and for subsequent figures. 

\begin{table}
\caption{Mapping Topological Features to Integer IDs for implementation }
\resizebox{0.5\textwidth}{!}{
\begin{tabular}{clc}
\toprule
 \textbf{Feature ID} & \textbf{Topological Feature Name} & \textbf{Homology Dimension} \\
\midrule
        0 & Bottleneck Amplitude & 0 \\
        1 &  Bottleneck Amplitude & 1 \\
\midrule
        2 &  Wasserstein Amplitude & 0 \\
        3 &  Wasserstein Amplitude & 1 \\
\midrule
        4 &  Betti Amplitude & 0 \\
        5 & Betti Amplitude & 1 \\
\midrule
        6 & Landscape Amplitude & 0 \\
        7 & Landscape Amplitude & 1 \\
\midrule
        8 & Silhouette Amplitude & 0 \\
        9 & Silhouette Amplitude & 1 \\
\midrule
        10 & Heat Amplitude & 0 \\
        11 & Heat Amplitude & 1 \\
\midrule
        12 & Normalized Persistence Entropy & 0 \\
        13 & Normalized Persistence Entropy & 1 \\
\midrule
        14 & Unnormalized Persistence Entropy & 0 \\
        15 & Unnormalized Persistence Entropy & 1 \\
\midrule
        16 & Number of Points & 0 \\
        17 & Number of Points & 1 \\
\bottomrule
\label{table:ID_to_Feature}
\end{tabular}}
\end{table}

\subsection{\textbf{Machine Learning Pipeline}} 

Once that both Power Band and Topological features have been extracted for each ECoG sample, we decided to stick to low-complexity classification models in a supervised machine learning pipeline. \\

Model selection has been carried out through a careful hyperparameter optimization procedure which is discussed in the next paragraphs.
Furthermore, we compared the performances of the algorithms we employed, by training them separately on Power Band features and on Topological features, and we showed that aggregating them considerably improves the performances in both cases. 

\paragraph{Model Selection.}

Specifically, we employed and compared the following classification algorithms: Random Forest, Gradient Boosting, Support Vector Machine, Multilayer Perceptron and Gaussian Naive Bayes. We decided not to utilize any deep learning architecture due to the scarcity of samples: in fact, we had only $180$ samples in total, where $90$ of them - i.e. 50\% - represented the Relax state, $30$ the Rock state, $30$ the Scissor state and  $30$ the Paper state.  \\

Given that our problem is multiclass and unbalanced, after hyperparameter optimization, we used 5-fold Stratified Cross Validation and selected the two best models accordingly: Random Forest and Gradient Boosting. Results from the chosen models are shown in \textit{Table \ref{table:hyperopt_rf_acc}} and \textit{Table \ref{table:hyperopt_gb_acc}}, while all the other tested models are shown in \textbf{Appendix}.

\paragraph{Hyperparameters Optimization.}

Machine learning models usually have several hyperparameters which can be tuned to change the way the learning process for that algorithm works, e.g. some of them act as regularizers. Modifying their values results in different predictive performance of the algorithm, which are likely tuned for a specific metric, i.e. accuracy in our case. Given that there is not a way to obtain universally optimal hyperparamters for the same algorithm on different data, we would like to find the best way - i.e. the fastest - in order to reach a reasonable accuracy score. \\ 

The simplest and most naive automated way is applying vanilla Grid Search: a completely deterministic procedure, where by specifying a range and a granularity for each hyperparameter, the algorithm tries all the possible combinations. Needless to say, the time complexity is exponential in the number of parameters, leading to infeasibility in most of the problems. 
Another automated method which works reasonably better is Randomized Grid Search, i.e. trying in a random fashion a combination of hyperparameters  and selecting the combination which returns the best result. 
However, there exist an even more efficient way to optimize the hyperparameters: Bayesian optimization \cite{10.5555/2999325.2999464}. \\

Within the Bayesian optimization framework, we utilize a \textit{surrogate} model to estimate the performance of our predictive algorithm as a function of hyperparameters values. This surrogate model is then used to select the next hyperparameter combination to try. In our specific case, we employed Gaussian Processes as the surrogate model, even though other surrogates might be employed, e.g. Random Forest, Tree-structured Parzen Estimator \cite{10.5555/2986459.2986743}. \\

Based on this method we selected the two most performing algorithms as classifiers: Random Forest and Gradient Boosting, see \textit{Figure \ref{fig:HyperOpt_RF_GB}} and \textit{Figure \ref{fig:HyperOpt_SVM_MLP_GNB}} in \textbf{Appendix} for convergence plots;  \textit{Table \ref{table:hyperopt_rf_acc}},  \textit{Table \ref{table:hyperopt_gb_acc}} and \textit{Table \ref{table:hyperopt_svm_acc}}, \textit{Table \ref{table:hyperopt_mlp_acc}}, \textit{Table \ref{table:hyperopt_gnb_acc}} in \textbf{Appendix} for hyperparameters values as well as accuracy results with respect to the four different preprocessing approaches (1 Hz CAR, 100 Hz CAR, 100 Hz, 50 Hz CAR).

\begin{figure*}
\centering
\includegraphics[width=\textwidth]{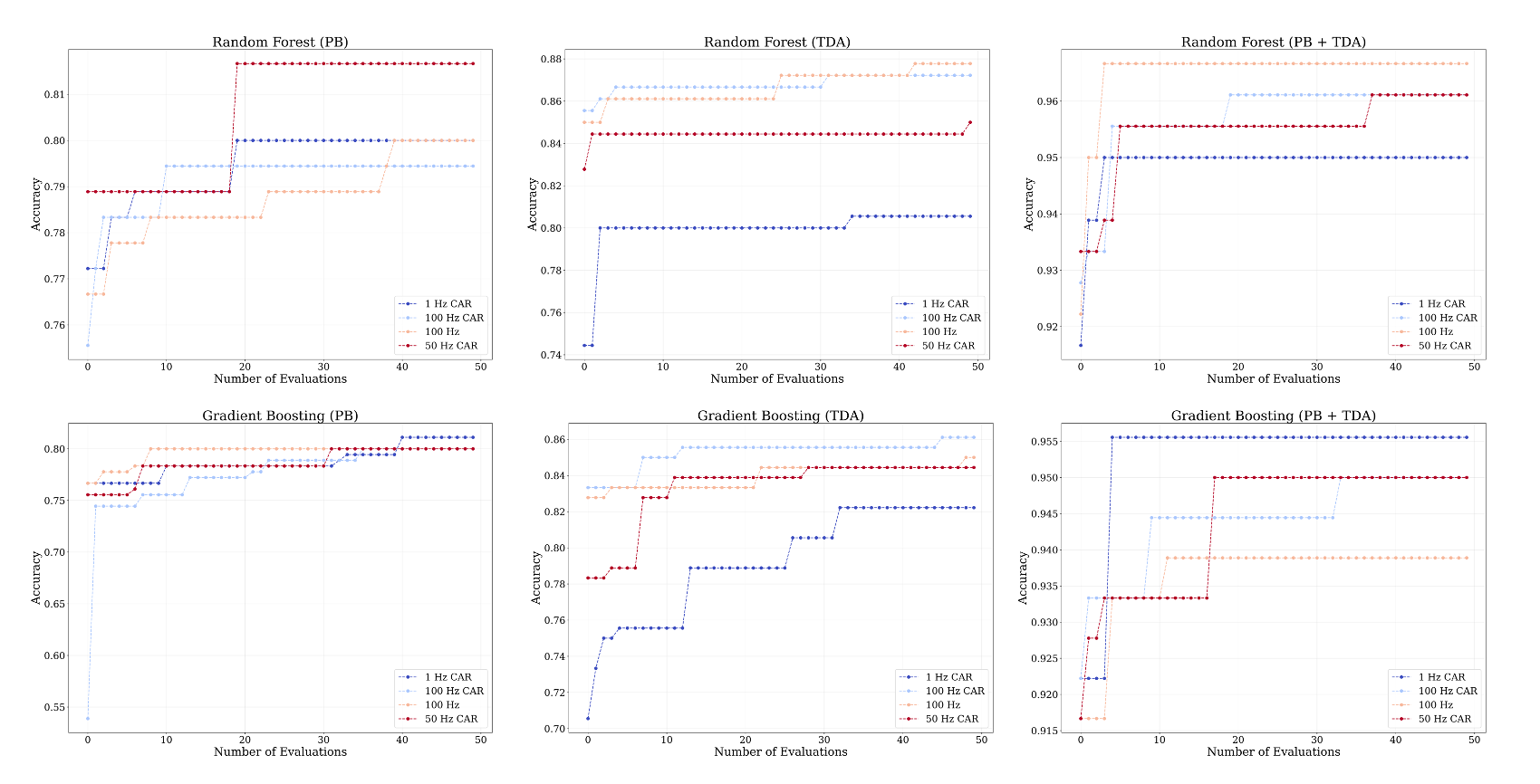}
\caption{Convergence plots for Hyperparameter Optimization through Gaussian Processes for Random Forest (Top) and Gradient Boosting (Bottom). From left to right: PB, Power Band Features; TDA, topological data analysis features; PB + TDA, power band and topological aggregated features.}
\label{fig:HyperOpt_RF_GB}
\end{figure*}

\begin{table}
\caption{Random Forest Hyperparameters Optimization and Results}
\resizebox{0.5\textwidth}{!}{
\begin{tabular}{lrrrrrr}
\toprule
 \textbf{Features} & max\_depth &  n\_estimators & criterion &  max\_features &  Accuracy &  Std dev \\
\midrule
      \textbf{Power Band} &   4 &           100 &      gini &         0.532 &      0.76 &     0.06 \\
         & 7 &           670 &   entropy &         0.049 &      0.78 &     0.03 \\
         & 8 &          1497 &      gini &         0.111 &      0.78 &     0.05 \\
         & 9 &           196 &   entropy &         0.066 &      0.78 &     0.04 \\
\midrule
       \textbf{TDA} &   8 &          1500 &   entropy &         1.000 &      0.80 &     0.05 \\
         &  8 &          1361 &   entropy &         0.131 &      0.86 &     0.06 \\
         &  3 &          1351 &   entropy &         1.000 &      0.87 &     0.05 \\
         &  4 &           217 &   entropy &         0.173 &      0.84 &     0.05 \\
         
\midrule
        \textbf{Power Band + TDA} &  6 &          1500 &   entropy &         0.240 &      0.94 &     0.04 \\
         &  6 &          1037 &      gini &         0.139 &      0.95 &     0.03 \\
         &  5 &           203 &   entropy &         0.106 &      0.95 &     0.04 \\
         &  5 &          1500 &   entropy &         0.174 &      0.96 &     0.03 \\
\bottomrule
\label{table:hyperopt_rf_acc}
\end{tabular}}
\end{table}

\begin{table}
\caption{Gradient Boosting Hyperparameters Optimization and Results}
\resizebox{0.5\textwidth}{!}{
\begin{tabular}{lrrrrrr}
\toprule
\textbf{Features} & max\_depth &  n\_estimators &    criterion &  subsample &  Accuracy &  Std Dev \\
\midrule
        \textbf{Power Band} &  3 &           100 &          mse &      0.474 &      0.77 &     0.06 \\
         & 3 &           100 &          mse &      0.555 &      0.77 &     0.06 \\
         & 4 &           554 &          mse &      0.544 &      0.78 &     0.06 \\
         & 3 &           891 & friedman\_mse &      0.567 &      0.78 &     0.05 \\
\midrule
        \textbf{TDA} & 6 &           100 &          mse &      1.000 &      0.81 &     0.03 \\
        & 10 &           100 &          mse &      0.853 &      0.85 &     0.04 \\
        & 10 &          1500 & friedman\_mse &      0.880 &      0.83 &     0.05 \\
        & 10 &           116 & friedman\_mse &      0.934 &      0.82 &     0.03 \\
\midrule
        \textbf{Power Band + TDA} & 8 &           446 & friedman\_mse &      0.721 &      0.94 &     0.05 \\
        & 10 &           886 &          mse &      0.835 &      0.92 &     0.06 \\
         & 4 &           451 &          mse &      0.399 &      0.92 &     0.06 \\
         & 6 &           700 &          mse &      0.901 &      0.93 &     0.03 \\
\bottomrule
\label{table:hyperopt_gb_acc}
\end{tabular}}

\end{table}

\paragraph{Feature Importance.}
When introducing a novel paradigm for feature extraction, i.e. topological data analysis, it is considered a good practice to quantify whether the information introduced is redundant or not with respect to classical techniques. \\

\begin{figure}[t]
\centering
\includegraphics[width=0.5\textwidth]{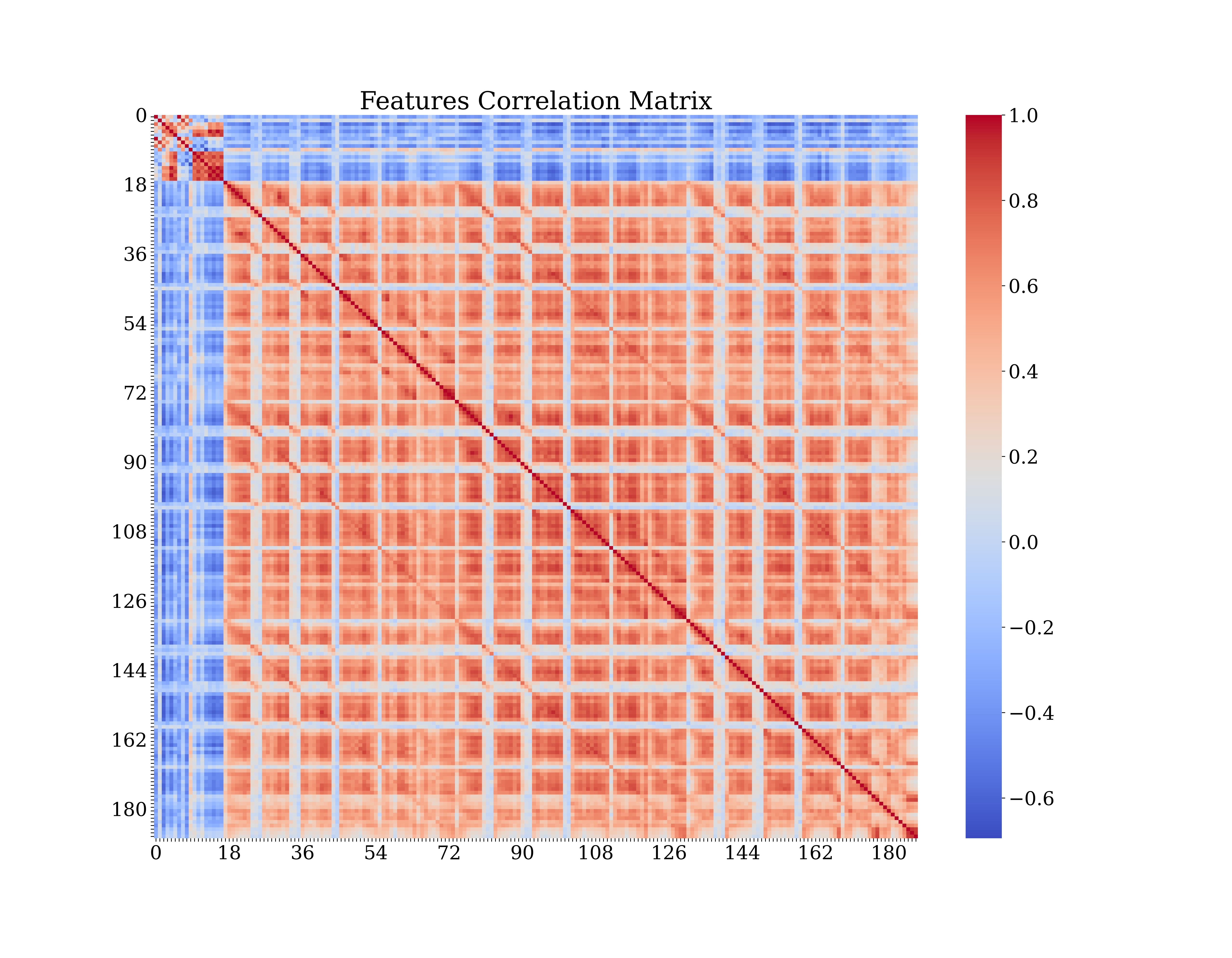}
\caption{Features Correlation Matrix. There's a clear distinction between the first 18 features (from 0 to 17), i.e. topological features, and power band features.}
\label{fig:corr_matrix}
\end{figure}

A possible model agnostic method is to consider the correlation matrix of features across samples. In \textbf{Figure \ref{fig:corr_matrix}}, it is immediate to notice that Topological features (i.e. from $0$ to $17$) are not at all correlated with Power Band features (i.e. from $18$ to $198$). Starting from this block diagonal structure, we further investigated the notion of Mutual Information between features and target variables, i.e. labels. 

\begin{figure}[t]
\centering
\includegraphics[width=0.45\textwidth]{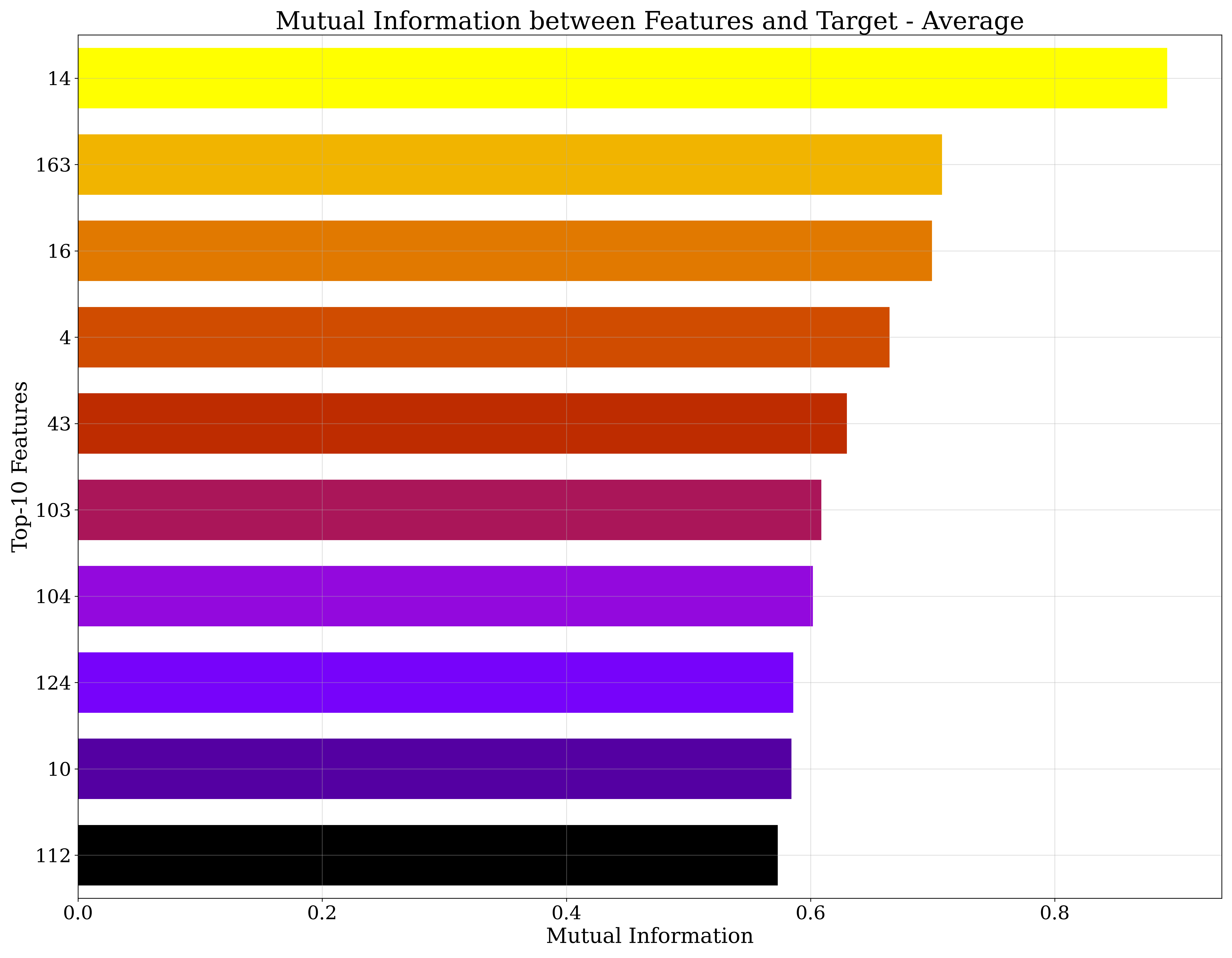}
\caption{Mutual Information between Top-10 Features and Target Variables, averaged on the four different preprocessed datasets. Features $14$, $16$, $4$ and $10$ are topological.}
\label{fig:MutualInforAvg}
\end{figure}

\begin{figure}[t]
\centering
\includegraphics[width=0.45\textwidth]{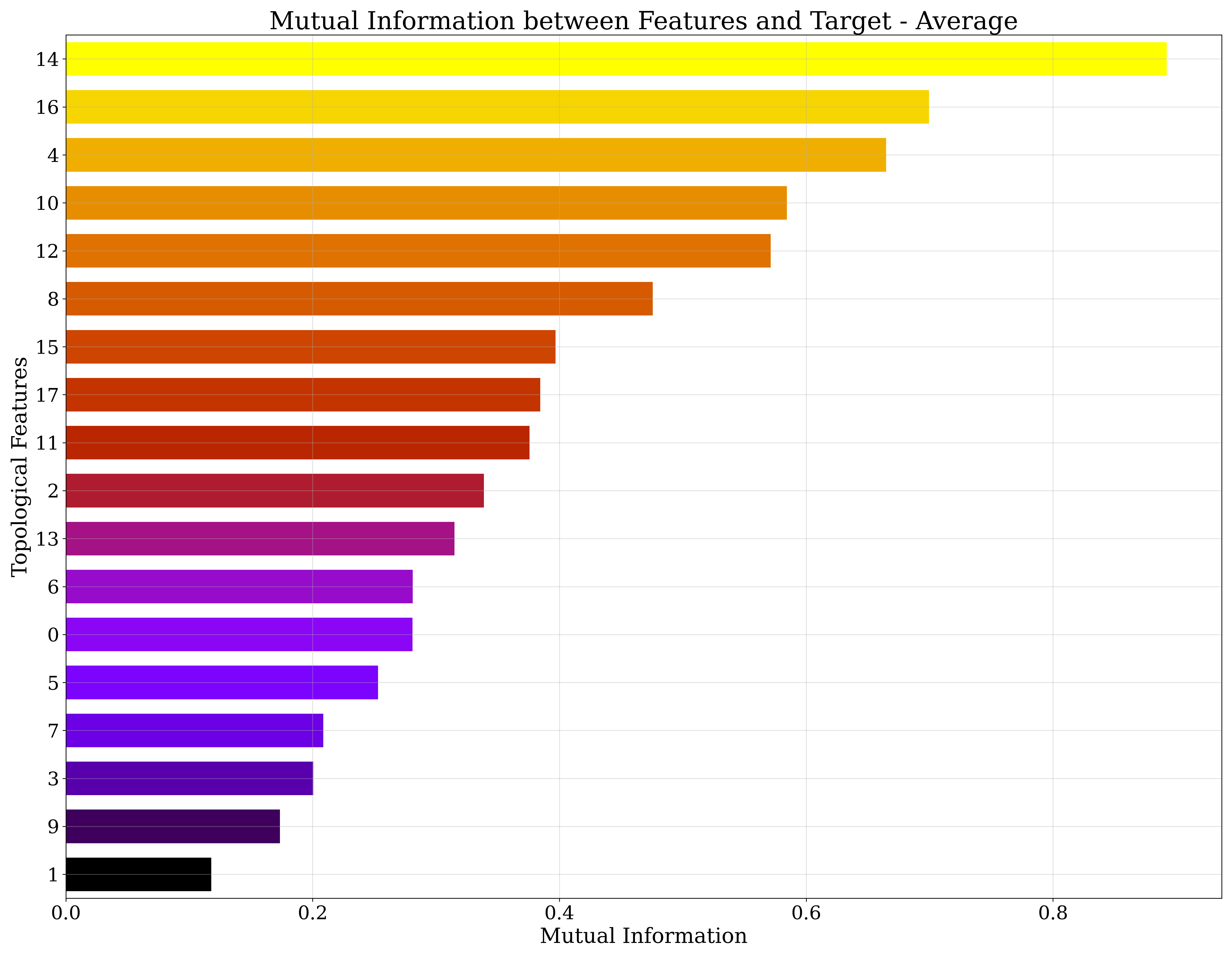}
\caption{Mutual Information between Topological Features and Target Variables, averaged on the four different preprocessed datasets}
\label{fig:MutualInfoTDAAvg}
\end{figure}

In particular, in \textit{Figure \ref{fig:MutualInforAvg}} are reported the $10$ most informative features, averaged on the four different preprocessing approaches: given that 4-out-of-10, as well as the first one, are Topological features, the impact is inevitably positive. As a comparative analysis, in \textit{Figure \ref{fig:MutualInfoTDAAvg}}, all the $18$ Topological features are shown in terms of Mutual Information, see also \textit{Table \ref{table:ID_to_Feature}} for TDA features to ID correspondence. \\

Multicollinearity is evident, due to the block diagonal structure of the correlation matrix. In this case, tackling feature importance with a permutation based approach \cite{breiman2001random} would not lead to a satisfying result. Since our two best classification models - i.e. Random Forest and Gradient Boosting - are both Tree-Based we opted for a model-specific approach based on node impurity \cite{breiman2017classification}. Node impurity is one of the intrinsic metrics used to split a branch in a Tree-based model, which basically disclose how much of a node belongs to a class. Given that feature importance is a relative measure - i.e. depends on data and specific model - and that we are considering four different preprocessing approaches and two classification models, a sharp way to compare results across different settings would be to consider features' rank.  Rank can be further averaged over the different preprocessing approaches and compared from model to model. In \textit{Table \ref{table:RF_feature_importance}} and \textit{Table \ref{table:GB_Feature_Importance}} we report feature importance and rank results for each preprocessing approach - e.g \textit{rf\_imp} stands for Random Forest feature specific importance on the first preprocessed dataset, \textit{rf\_imp\_1} on the second preprocessed dataset and so on. As we have foreseen with mutual information, topological features are relevant for both Random Forest and Gradient Boosting classifiers. \\ 

The proposed feature importance approach shed light on the relevance of topological features and can be seen an insightful perspective on specific ones, such as feature $16$, $14$ and $4$, respectively \textit{Number of Points} (first dimension), \textit{Persistence Entropy} (first dimension) and \textit{Amplitude} with \textit{Betti} metric (first dimension). \\

Different approaches for comparing features' importance exist: for example, instead of considering the rank, we could have scaled importance measures so that their lowest values are $0$ and highest are $1$ - i.e. min-max scaling. In the latter case, we might have ended up revealing information about the relative spacing in importance between features and less about their order.


\begin{table*}
\caption{Random Forest Top 10 Feature Importance}
\resizebox{\textwidth}{!}{
\begin{tabular}{rrrrrrrrrrr}
\toprule
Feature ID &   rf\_imp &  rf\_rank &  rf\_imp\_1 &  rf\_rank\_1 &  rf\_imp\_2 &  rf\_rank\_2 &  rf\_imp\_3 &  rf\_rank\_3 &  avg\_imp &  avg\_rank \\
\midrule
   16 & 0.178636 &        1 &  0.110822 &          2 &  0.117674 &          1 &  0.114373 &          2 & 0.130376 &      1.50 \\
   14 & 0.127132 &        2 &  0.110936 &          1 &  0.113162 &          2 &  0.121215 &          1 & 0.118111 &      1.50 \\
    4 & 0.029101 &        6 &  0.055215 &          3 &  0.056537 &          3 &  0.051734 &          3 & 0.048147 &      3.75 \\
  103 & 0.095553 &        3 &  0.047188 &          4 &  0.046532 &          4 &  0.043499 &          5 & 0.058193 &      4.00 \\
  163 & 0.095164 &        4 &  0.043100 &          5 &  0.043001 &          5 &  0.048162 &          4 & 0.057357 &      4.50 \\
   43 & 0.051568 &        5 &  0.033094 &          6 &  0.030807 &          8 &  0.033556 &          7 & 0.037256 &      6.50 \\
  112 & 0.025804 &        7 &  0.021518 &          9 &  0.014634 &         13 &  0.021236 &          9 & 0.020798 &      9.50 \\
   10 & 0.008552 &       18 &  0.026356 &          8 &  0.033624 &          7 &  0.026457 &          8 & 0.023747 &     10.25 \\
  122 & 0.017238 &       10 &  0.016938 &         12 &  0.017279 &         11 &  0.017943 &         10 & 0.017350 &     10.75 \\
  123 & 0.023991 &        8 &  0.020730 &         11 &  0.016947 &         12 &  0.016049 &         13 & 0.019429 &     11.00 \\
\bottomrule
\label{table:RF_feature_importance}
\end{tabular}}
\end{table*}

\begin{table*}
\caption{Gradient Boosting Top 10 Feature Importance}
\resizebox{\textwidth}{!}{
\begin{tabular}{rrrrrrrrrrr}
\toprule
Feature ID &   gb\_imp &  gb\_rank &  gb\_imp\_1 &  gb\_rank\_1 &  gb\_imp\_2 &  gb\_rank\_2 &  gb\_imp\_3 &  gb\_rank\_3 &  avg\_imp &  avg\_rank \\
\midrule
   16 & 0.261595 &        1 &  0.336517 &          1 &  0.178810 &          2 &  0.374422 &          1 & 0.287836 &      1.25 \\
  122 & 0.133643 &        3 &  0.137376 &          2 &  0.058210 &          3 &  0.143441 &          2 & 0.118168 &      2.50 \\
  103 & 0.090184 &        4 &  0.100476 &          3 &  0.049026 &          4 &  0.072926 &          4 & 0.078153 &      3.75 \\
   14 & 0.152591 &        2 &  0.041389 &          6 &  0.204489 &          1 &  0.020085 &          9 & 0.104638 &      4.50 \\
  154 & 0.028249 &        8 &  0.039477 &          7 &  0.032239 &          7 &  0.046630 &          5 & 0.036649 &      6.75 \\
    4 & 0.034076 &        6 &  0.050935 &          5 &  0.017181 &         13 &  0.022776 &          8 & 0.031242 &      8.00 \\
   10 & 0.033223 &        7 &  0.058431 &          4 &  0.009667 &         21 &  0.086715 &          3 & 0.047009 &      8.75 \\
  124 & 0.049691 &        5 &  0.005141 &         21 &  0.021571 &         11 &  0.043781 &          6 & 0.030046 &     10.75 \\
   43 & 0.012844 &       11 &  0.004685 &         22 &  0.048069 &          5 &  0.003847 &         23 & 0.017361 &     15.25 \\
   47 & 0.011553 &       13 &  0.015106 &          9 &  0.002932 &         41 &  0.016968 &         10 & 0.011640 &     18.25 \\
\bottomrule
\label{table:GB_Feature_Importance}
\end{tabular}}
\end{table*}


\section{\textbf{DISCUSSION}}\label{discussion}

\subsection{\textbf{Discussion of the experimental results}}
In this section we summarize and comment the experimental results obtained using different classification models.
\newline
\newline
We start by pointing out the importance and the value of the topological features.
Fig. \ref{fig:corr_matrix} shows the features correlation matrix, and it clearly indicated that TDA and PB features are not correlated. This means that they contain different information that could be useful to use together.
The importance of the TDA features is also supported by the feature importance analysis reported in Table \ref{table:RF_feature_importance} and Table \ref{table:GB_Feature_Importance} and in Fig. \ref{fig:MutualInfoTDAAvg}. In both the models used, topological features appeared among the most informative ones, based on mutual information with the target variables. This is remarkable, considering that topological features are outnumbered by power band ones.
Finally in Table \ref{table:hyperopt_rf_acc} and Table \ref{table:hyperopt_gb_acc}, and in Figure \ref{fig:HyperOpt_RF_GB}, it is possible to see the difference in the performance of the models, by using the two kinds of features alone or aggregated.
In both selected models we saw that the usage of PB and TDA features combined leads to the best accuracy. This indicates that the two kind of features represents information that are complementary for the problem of classification. 
\newline
Established the importance of TDA features we focused on model selection and hyperparameter optimization.
Tables \ref{table:hyperopt_rf_acc}, \ref{table:hyperopt_gb_acc}, \ref{table:hyperopt_svm_acc}, \ref{table:hyperopt_mlp_acc}, \ref{table:hyperopt_gnb_acc}, indicate that the models with the best performance are Random Forest and Gradient boosting. 
Among those two, the best results in terms of accuracy are obtained in random forest. With this models it is possible to obtain an accuracy of 0.96, after an accurate hyperparameter optimization. This value of accuracy is obtained from stratified cross validation, and is compatible with the state of the art \cite{gruenwald2019time}.

\subsection{\textbf{Advantages of this technique}}\label{advantages}

A key component of the proposed method is the usage of TDA as a complementary framework to classical BCI techniques. Such an approach exploits the geometrical properties of the data to infer a particular kind of features with a wide range of advantages.\\

Firs of all, features extracted via PH are robust to noise and small perturbation of the input data.
Moreover, they are generally invariant for translation, rotation and reflection of the input.\\

Secondly, they contain information that are different from the ones encoded in the power band features, as shown in \textbf{Figure \ref{fig:corr_matrix}}. Such a difference indicates that the topological approach improves our knowledge of the input data and provides additional information to be exploited during the classification phase. 


\subsection{\textbf{Limitations}}

A critical point of our work is the reduced number of ECoG samples we used to carry out the analysis. A single patient recording was in fact considered.\\

Moreover, we only focused on a specific four-class classification problem. Further analysis should be made to confirm the validity of our method to a wider class of motor experiments and to a different number of classes.

\section{\textbf{FUTURE WORK}}

The research proposed in this paper is the first, to the best of our knowledge, concrete application of TDA on ECoG neural recordings. Given its prematurity, we would like to further deepen the following aspects:

\begin{itemize}
    \itemsep0em
    \item In the TDA feature extraction part, we exploited only low-order homology dimensions: higher dimensionality could be considered and compared with the current ones; 
    \item Another interesting perspective is understanding whether it is possible to make our pipeline near-real-time for on-the-field applications, for instance GPU implementations of TDA's techniques - i.e. through \textit{giotto-ph} \cite{DBLP:journals/corr/abs-2107-05412}- or considering Persistent Cohomology based techniques \cite{deSilvaVin2011Dip};  ; 
    \item Further investigate partial dependence among topological features and power band features, i.e. a reverse-engineering approach might carry some discoveries with respect to the role of specific topological features; 
    \item Try to generalize the same approach on different data: a larger number of patients,  different kind of recording (e.g. EEG) or even other types of time series, in order to attempt to set a standardized pipeline and discern in which cases results are relevant;
    \item Apply channel exclusion techniques to select the best channels and avoid redundant and noisy information.
\end{itemize}

\section{\textbf{HARDWARE SUPPORT \& CODE}}

We acknowledge the University of Turin's and Polytechnic University of Turin's High Performance Centre for Artificial Intelligence (\textbf{HPC4AI}) for providing us with the following computational resources: 

\begin{itemize}
    \itemsep0em
    \item CPU: 12 vCPU Intel Xeon
    \item RAM: 20 GB RAM ECC
    \item OS: Ubuntu Server 20.04 LTS
    \item Python version: 3.7.3
    \item GPU: Nvidia Tesla T4 (not used yet)
\end{itemize}

Moreover, the implementation described in the previous paragraphs is largely based on the following libraries: TDA's feature extraction with
\textit{giotto-tda} \cite{tauzin2020giottotda}, hyperparameter optimization \textit{scikit-optimize} \cite{tim_head_2018_1207017} along with Machine Learning algorithms implementation with \textit{scikit-learn} \cite{scikit-learn}. 

All the code produced is published on a public repository available at:  \url{https://github.com/MachineLearningJournalClub/ECoG_VBH_2021}.\\


\section{\textbf{CONCLUSIONS}}
The study we carried out had the goal to investigate the applicability of the recently developed field of Topological Data Analysis on time series obtained from neural recordings.
We obtained solid evidence that topological features extracted with persistence homology from ECoG data are informative and complementary to standard features as power band. Using both kind of features we observed a significant improvement in classification accuracy of data from a single patient performing hand gestures.
More work is needed to validate the efficacy of our approach and to extend it to other cases of study.  On the other side, results show potential for attempting to generalize it on different task of classification of neural recordings.
The encouraging outcome inspires us to continue our work by testing the methods on more data and possibly devise a standardized TDA pipeline for such a kind of time series.

\section{\textbf{ACKNOWLEDGEMENTS}}
This work was presented at the \textit{Virtual Brain Hackathon} organized by \textbf{g.tec} in April 2021.
We acknowledge University of Turin, Polytechnic University of Turin, Machine Learning Journal Club, HPC4AI for supporting us.

\clearpage 

\printbibliography

@article{art1,
  title={Choosing your reference--and why it matters},
  author={Leuchs, Laura},
  journal={Brain Products},
  pages={03--05},
  year={2019}
}

@article{Lotte_2018,
	author={F Lotte and L Bougrain and A Cichocki and M Clerc and M Congedo and A Rakotomamonjy and F Yger},
	title={A review of classification algorithms for {EEG}-based brain{\textendash}computer interfaces: a 10 year update},
	journal={Journal of Neural Engineering},
	year={2018},
	month={apr},
	publisher={{IOP} Publishing},
	volume={15},
	number={3},
	pages={031005}
}

@article{Bout_2016,
    author={Bouton C., Shaikhouni A., Annetta N. et al.},
    title={Restoring cortical control of functional movement in a human with quadriplegia},
    journal={Nature},
    year={2016},
    volume={533},
    pages={247-250}
}

@book{hari2017meg,
  title={MEG-EEG Primer},
  author={Hari, R. and Puce, A.},
  isbn={9780190497774},
  lccn={2016039489},
  year={2017},
  publisher={Oxford University Press}
}

@article{sig_process,
author = {Lotte, Fabien},
year = {2014},
month = {10},
pages = {},
title = {A Tutorial on EEG Signal Processing Techniques for Mental State Recognition in Brain-Computer Interfaces},
}

@article{6165246,
  author={van Erp, Jan and Lotte, Fabien and Tangermann, Michael},
  journal={Computer}, 
  title={Brain-Computer Interfaces: Beyond Medical Applications}, 
  year={2012},
  volume={45},
  number={4},
  pages={26-34}
}

@article{Chestek_2013,
    author={Cynthia A Chestek and Vikash Gilja and Christine H Blabe and Brett L Foster and Krishna V Shenoy and Josef Parvizi and Jaimie M Henderson},
    journal={Journal of Neural Engineering},
    title={Hand posture classification using electrocorticography signals in the gamma band over human sensorimotor brain areas},
	year={2013},
	publisher = {{IOP} Publishing},
	volume = {10},
	number = {2},
	pages = {026002},
}

@article{gruenwald2019time,
  title={Time-variant linear discriminant analysis improves hand gesture and finger movement decoding for invasive brain-computer interfaces},
  author={Gruenwald, Johannes and Znobishchev, Andrei and Kapeller, Christoph and Kamada, Kyousuke and Scharinger, Josef and Guger, Christoph},
  journal={Frontiers in neuroscience},
  volume={13},
  pages={901},
  year={2019},
  publisher={Frontiers}
}

@article{jiang2020power,
  title={Power modulations of ecog alpha/beta and gamma bands correlate with time-derivative of force during hand grasp},
  author={Jiang, Tianxiao and Pellizzer, Giuseppe and Asman, Priscella and Bastos, Dhiego and Bhavsar, Shreyas and Tummala, Sudhakar and Prabhu, Sujit and Ince, Nuri F},
  journal={Frontiers in neuroscience},
  volume={14},
  pages={100},
  year={2020},
  publisher={Frontiers}
}

@article{li2017gesture,
  title={Gesture decoding using ECoG signals from human sensorimotor cortex: a pilot study},
  author={Li, Yue and Zhang, Shaomin and Jin, Yile and Cai, Bangyu and Controzzi, Marco and Zhu, Junming and Zhang, Jianmin and Zheng, Xiaoxiang},
  journal={Behavioural neurology},
  volume={2017},
  year={2017},
  publisher={Hindawi}
}

@article{Taku2011electro,
  title={Real-time control of a prosthetic hand using human electrocorticography signals},
  author={Yanagisawa and Masayuki Hirata and Youichi Saitoh and Tetsu Goto and Haruhiko Kishima and Ryohei Fukuma and Hiroshi Yokoi and Yukiyasu Kamitani and Toshiki Yoshimine},
  journal={Journal of Neurosurgery JNS},
  volume={114},
  number={6},
  pages={1715 - 1722},
  year={2011},
  publisher={American Association of Neurological Surgeons},
}

@article{staba2002quantitative,
  title={Quantitative analysis of high-frequency oscillations (80--500 Hz) recorded in human epileptic hippocampus and entorhinal cortex},
  author={Staba, Richard J and Wilson, Charles L and Bragin, Anatol and Fried, Itzhak and Engel Jr, Jerome},
  journal={Journal of neurophysiology},
  volume={88},
  number={4},
  pages={1743--1752},
  year={2002},
  publisher={American Physiological Society Bethesda, MD}
}

@inproceedings{kapeller2014single,
  title={Single trial detection of hand poses in human ECoG using CSP based feature extraction},
  author={Kapeller, Christoph and Schneider, Christoph and Kamada, Kyousuke and Ogawa, Hiroshi and Kunii, Naoto and Ortner, Rupert and Prueckl, Robert and Guger, Christoph},
  booktitle={2014 36th Annual International Conference of the IEEE Engineering in Medicine and Biology Society},
  pages={4599--4602},
  year={2014},
  organization={IEEE}
}

@article{WassermanLarry2018TDA,
author = {Wasserman, Larry},
copyright = {Copyright © 2018 by Annual Reviews. All rights reserved 2018},
issn = {2326-8298},
journal = {Annual review of statistics and its application},
language = {eng},
number = {1},
pages = {501-532},
publisher = {Annual Reviews},
title = {Topological Data Analysis},
volume = {5},
year = {2018},
}

@MISC{Edelsbrunner_persistenthomology,
    author = {Herbert Edelsbrunner and John Harer},
    title = { Persistent Homology -- a Survey},
    year = {}
}

@book{edelsbrunner2010computational,
  title={Computational topology: an introduction},
  author={Edelsbrunner, Herbert and Harer, John},
  year={2010},
  publisher={American Mathematical Soc.}
}

@article{WuChengyuan2021Tmlf,
author = {Wu, Chengyuan and Hargreaves, Carol Anne},
issn = {0952-813X},
journal = {Journal of experimental \& theoretical artificial intelligence},
language = {eng},
pages = {1-16},
title = {Topological machine learning for multivariate time series},
year = {2021},
}

@inproceedings{garin2019topological,
  title={A topological" reading" lesson: Classification of MNIST using TDA},
  author={Garin, Ad{\'e}lie and Tauzin, Guillaume},
  booktitle={2019 18th IEEE International Conference On Machine Learning And Applications (ICMLA)},
  pages={1551--1556},
  year={2019},
  organization={IEEE}
}

@article{BiasottiS2008Dsbg,
author = {Biasotti, S and De Floriani, L and Falcidieno, B and Frosini, P and Giorgi, D and Landi, C and Papaleo, L and Spagnuolo, M},
address = {New York, NY},
copyright = {2009 INIST-CNRS},
issn = {0360-0300},
journal = {ACM computing surveys},
language = {eng},
number = {4},
pages = {1-87},
publisher = {ACM},
title = {Describing shapes by geometrical-topological properties of real functions},
volume = {40},
year = {2008},
}

@article{CarlssonGunnar2014Tprf,
author = {Carlsson, Gunnar},
address = {Cambridge, UK},
copyright = {Copyright © Cambridge University Press 2014},
issn = {0962-4929},
journal = {Acta numerica},
language = {eng},
pages = {289-368},
publisher = {Cambridge University Press},
title = {Topological pattern recognition for point cloud data},
volume = {23},
year = {2014},
}

@article{PanGang2018RDoH,
author = {Pan, Gang and Li, Jia-Jun and Qi, Yu and Yu, Hang and Zhu, Jun-Ming and Zheng, Xiao-Xiang and Wang, Yue-Ming and Zhang, Shao-Min},
address = {Switzerland},
copyright = {COPYRIGHT 2018 Frontiers Research Foundation},
issn = {1662-4548},
journal = {Frontiers in neuroscience},
language = {eng},
pages = {555-555},
publisher = {Frontiers Research Foundation},
title = {Rapid Decoding of Hand Gestures in Electrocorticography Using Recurrent Neural Networks},
volume = {12},
year = {2018},
}

@article{LiYue2017GDUE,
author = {Li, Yue and Zhang, Shaomin and Jin, Yile and Cai, Bangyu and Controzzi, Marco and Zhu, Junming and Zhang, Jianmin and Zheng, Xiaoxiang},
address = {Netherlands},
copyright = {Copyright © 2017 Yue Li et al.},
issn = {0953-4180},
journal = {Behavioural neurology},
language = {eng},
pages = {3435686-12},
publisher = {Hindawi},
title = {Gesture Decoding Using ECoG Signals from Human Sensorimotor Cortex: A Pilot Study},
volume = {2017},
year = {2017},
}

@inproceedings{10.5555/2986459.2986743,
author = {Bergstra, James and Bardenet, R\'{e}mi and Bengio, Yoshua and K\'{e}gl, Bal\'{a}zs},
title = {Algorithms for Hyper-Parameter Optimization},
year = {2011},
isbn = {9781618395993},
publisher = {Curran Associates Inc.},
address = {Red Hook, NY, USA},
booktitle = {Proceedings of the 24th International Conference on Neural Information Processing Systems},
pages = {2546–2554},
numpages = {9},
location = {Granada, Spain},
series = {NIPS'11}
}

@inproceedings{10.5555/2999325.2999464,
author = {Snoek, Jasper and Larochelle, Hugo and Adams, Ryan P.},
title = {Practical Bayesian Optimization of Machine Learning Algorithms},
year = {2012},
publisher = {Curran Associates Inc.},
address = {Red Hook, NY, USA},
booktitle = {Proceedings of the 25th International Conference on Neural Information Processing Systems - Volume 2},
pages = {2951–2959},
numpages = {9},
location = {Lake Tahoe, Nevada},
series = {NIPS'12}
}

@book{breiman2017classification,
  title={Classification and regression trees},
  author={Breiman, Leo and Friedman, Jerome H and Olshen, Richard A and Stone, Charles J},
  year={2017},
  publisher={Routledge}
}

@article{breiman2001random,
  title={Random forests},
  author={Breiman, Leo},
  journal={Machine learning},
  volume={45},
  number={1},
  pages={5--32},
  year={2001},
  publisher={Springer}
}

@article{giotto-tda,
  author  = {Guillaume Tauzin and Umberto Lupo and Lewis Tunstall and Julian Burella P\'{e}rez and Matteo Caorsi and Anibal M. Medina-Mardones and Alberto Dassatti and Kathryn Hess},
  title   = {giotto-tda: A Topological Data Analysis Toolkit for Machine Learning and Data Exploration},
  journal = {Journal of Machine Learning Research},
  year    = {2021},
  volume  = {22},
  number  = {39},
  pages   = {1-6},
  url     = {http://jmlr.org/papers/v22/20-325.html}
}

@software{tim_head_2018_1207017,
  author       = {Tim Head and
                  MechCoder and
                  Gilles Louppe and
                  Iaroslav Shcherbatyi and
                  fcharras and
                  Zé Vinícius and
                  cmmalone and
                  Christopher Schröder and
                  nel215 and
                  Nuno Campos and
                  Todd Young and
                  Stefano Cereda and
                  Thomas Fan and
                  rene-rex and
                  Kejia (KJ) Shi and
                  Justus Schwabedal and
                  carlosdanielcsantos and
                  Hvass-Labs and
                  Mikhail Pak and
                  SoManyUsernamesTaken and
                  Fred Callaway and
                  Loïc Estève and
                  Lilian Besson and
                  Mehdi Cherti and
                  Karlson Pfannschmidt and
                  Fabian Linzberger and
                  Christophe Cauet and
                  Anna Gut and
                  Andreas Mueller and
                  Alexander Fabisch},
  title        = {scikit-optimize/scikit-optimize: v0.5.2},
  month        = mar,
  year         = 2018,
  publisher    = {Zenodo},
  version      = {v0.5.2},
  doi          = {10.5281/zenodo.1207017},
  url          = {https://doi.org/10.5281/zenodo.1207017}
}

@article{scikit-learn,
 title={Scikit-learn: Machine Learning in {P}ython},
 author={Pedregosa, F. and Varoquaux, G. and Gramfort, A. and Michel, V.
         and Thirion, B. and Grisel, O. and Blondel, M. and Prettenhofer, P.
         and Weiss, R. and Dubourg, V. and Vanderplas, J. and Passos, A. and
         Cournapeau, D. and Brucher, M. and Perrot, M. and Duchesnay, E.},
 journal={Journal of Machine Learning Research},
 volume={12},
 pages={2825--2830},
 year={2011}
}

@article{DBLP:journals/corr/abs-2107-05412,
  author    = {Julian Burella P{\'{e}}rez and
               Sydney Hauke and
               Umberto Lupo and
               Matteo Caorsi and
               Alberto Dassatti},
  title     = {Giotto-ph: {A} Python Library for High-Performance Computation of
               Persistent Homology of Vietoris-Rips Filtrations},
  journal   = {CoRR},
  volume    = {abs/2107.05412},
  year      = {2021},
  url       = {https://arxiv.org/abs/2107.05412},
  eprinttype = {arXiv},
  eprint    = {2107.05412},
  bibsource = {dblp computer science bibliography, https://dblp.org}
}

@misc{tauzin2020giottotda,
      title={giotto-tda: A Topological Data Analysis Toolkit for Machine Learning and Data Exploration},
      author={Guillaume Tauzin and Umberto Lupo and Lewis Tunstall and Julian Burella Pérez and Matteo Caorsi and Anibal Medina-Mardones and Alberto Dassatti and Kathryn Hess},
      year={2020},
      eprint={2004.02551},
      archivePrefix={arXiv},
      primaryClass={cs.LG}
}

@article{deSilvaVin2011Dip,
author = {de Silva, Vin and Morozov, Dmitriy and Vejdemo-Johansson, Mikael},
address = {United States},
issn = {0266-5611},
journal = {Inverse problems},
language = {eng},
number = {12},
organization = {Lawrence Berkeley National Lab. (LBNL), Berkeley, CA (United States)},
pages = {124003-},
publisher = {IOP Publishing},
title = {Dualities in persistent (co)homology},
volume = {27},
year = {2011},
}

\clearpage 

\section*{\textbf{APPENDIX}}\label{appendix}

Random Forest and Gradient Boosting are the most accurate models that we employed. On the other hand, during the model selection, we trained several other models that are reported here in \textbf{Appendix} for completeness. In particular, we report hyperparameters for: Support Vector Machine in \textit{Table \ref{table:hyperopt_svm_acc}}; Multilayer Perceptron in \textit{Table \ref{table:hyperopt_mlp_acc}}; and Gaussian Naive Bayes in \textit{Table \ref{table:hyperopt_gnb_acc}}.\\

\begin{figure*}
\centering
\includegraphics[width=\textwidth]{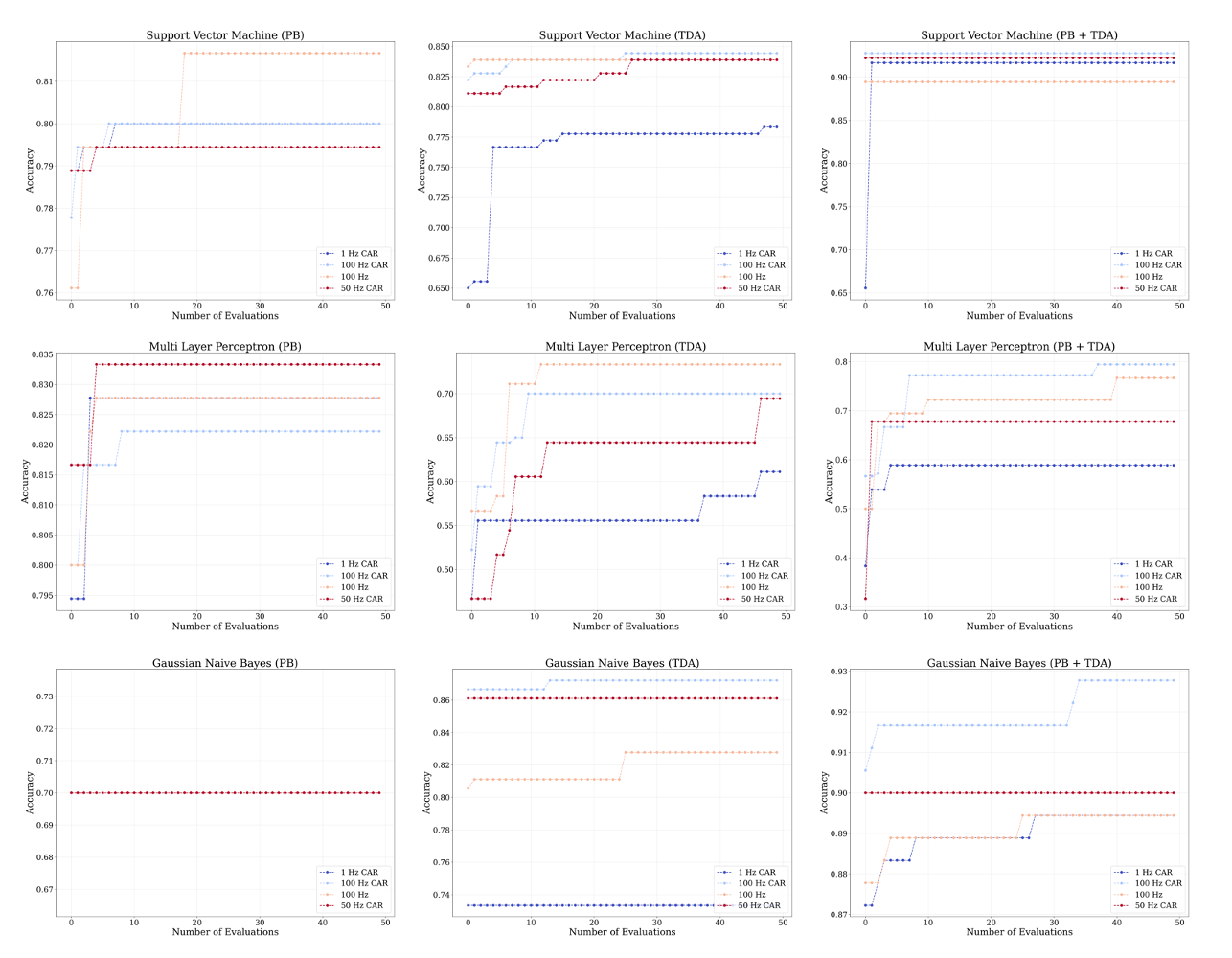}
\caption{Convergence plots for Hyperparameter Optimization through Gaussian Processes for Support Vector Machine (Top), Multi-Layer Perceptron (Middle) and Gaussian Naive Bayes (Bottom). From left to right: PB, Power Band Features; TDA, topological data analysis features; PB + TDA, power band and topological aggregated features.}
\label{fig:HyperOpt_SVM_MLP_GNB}
\end{figure*}

Convergence plots for hyperparameter optimization are also shown in \textit{Figure \ref{fig:HyperOpt_SVM_MLP_GNB}}.\\

\begin{table}[H]
\caption{Support Vector Machine Hyperparameters Optimization and Results}
\resizebox{0.5\textwidth}{!}{
\begin{tabular}{lrrrrrr}
\toprule
\textbf{Features} & C & kernel &  Accuracy &  Std Dev  \\
\midrule
        \textbf{Power Band} & 0.258 & linear &      0.80 &     0.03 \\
                            & 0.152 & linear &      0.80 &     0.01 \\
                            & 0.228 & linear &      0.82 &     0.01 \\
                             & 7.204 &    rbf &      0.79 &     0.03 \\
\midrule
        \textbf{TDA} & 2.263 & linear &      0.78 &     0.03 \\
                     & 10.000 & linear &      0.84 &     0.06 \\
                     & 8.255 &    rbf &      0.84 &     0.05 \\
                     & 0.001 & linear &      0.84 &     0.06 \\
\midrule
        \textbf{Power Band + TDA} & 3.182 & linear &      0.92 &     0.05 \\
                                  & 7.047 & linear &      0.93 &     0.04 \\
                                  & 5.709 & linear &      0.89 &     0.05 \\
                                  & 9.308 & linear &      0.92 &     0.03 \\
\bottomrule
\label{table:hyperopt_svm_acc}
\end{tabular}}
\end{table}

\begin{table}[H]
\caption{Multilayer Perceptron Hyperparameters Optimization and Results}
\resizebox{0.5\textwidth}{!}{
\begin{tabular}{lrrrrrr}
\toprule
\textbf{Features} & alpha &  Accuracy &  Std Dev  \\
\midrule
        \textbf{Power Band} & 0.054 &      0.79 &     0.05 \\
                            & 0.069 &      0.79 &     0.04 \\
                            & 0.043 &      0.79 &     0.05 \\
                            & 0.002 &      0.79 &     0.05 \\
\midrule
        \textbf{TDA}  & 0.007 &      0.45 &     0.14 \\
                      & 0.021 &      0.47 &     0.17 \\
                      & 0.008 &      0.52 &     0.15 \\
                      & 0.100 &      0.44 &     0.16 \\
\midrule
        \textbf{Power Band + TDA} & 0.040 &      0.46 &     0.13 \\
                                  & 0.100 &      0.59 &     0.17 \\
                                  & 0.077 &      0.58 &     0.20 \\
                                  & 0.089 &      0.50 &     0.18 \\
        
\bottomrule
\label{table:hyperopt_mlp_acc}
\end{tabular}}
\end{table}

\begin{table}[H]
\caption{Gaussian Naive Bayes Hyperparameters Optimization and Results}
\resizebox{0.5\textwidth}{!}{
\begin{tabular}{lrrrrrr}
\toprule
\textbf{Features} & var\_smoothing &  Accuracy &  Std Dev  \\
\midrule
        \textbf{Power Band} &  0.0 &       0.7 &     0.09 \\
          & 0.0 &       0.7 &     0.09 \\
          & 0.0 &       0.7 &     0.09 \\
          & 0.0 &       0.7 &     0.09 \\
\midrule
        \textbf{TDA}   & 5.055780e-08 &      0.73 &     0.07 \\
                       & 7.497147e-09 &      0.87 &     0.07 \\
                       & 1.289194e-10 &      0.83 &     0.06 \\
                       & 8.680831e-08 &      0.86 &     0.02 \\
\midrule
        \textbf{Power Band + TDA} & 3.564985e-08 &      0.89 &     0.08 \\
                                  & 1.000000e-10 &      0.93 &     0.05 \\
                                  & 1.231941e-10 &      0.89 &     0.10 \\
                                  &  1.449855e-08 &      0.90 &     0.06 \\
        
\bottomrule
\label{table:hyperopt_gnb_acc}
\end{tabular}}
\end{table}

Regarding feature importance, in \textit{Table \ref{table:TDA_RF_Feature_Imp}} and \textit{Table \ref{table:TDA_GB_Feature_Imp}} we show a full report for Topological Data Analysis features with Random Forest and Gradient Boosting. 

\begin{table*}
\caption{Random Forest TDA Feature Importance}
\resizebox{\textwidth}{!}{
\begin{tabular}{rrrrrrrrrrr}
\toprule
 name &   rf\_imp &  rf\_rank &  rf\_imp\_1 &  rf\_rank\_1 &  rf\_imp\_2 &  rf\_rank\_2 &  rf\_imp\_3 &  rf\_rank\_3 &  avg\_imp &  avg\_rank \\
\midrule
   14 & 0.127132 &        2 &  0.110936 &          1 &  0.113162 &          2 &  0.121215 &          1 & 0.118111 &      1.50 \\
   16 & 0.178636 &        1 &  0.110822 &          2 &  0.117674 &          1 &  0.114373 &          2 & 0.130376 &      1.50 \\
    4 & 0.029101 &        6 &  0.055215 &          3 &  0.056537 &          3 &  0.051734 &          3 & 0.048147 &      3.75 \\
   10 & 0.008552 &       18 &  0.026356 &          8 &  0.033624 &          7 &  0.026457 &          8 & 0.023747 &     10.25 \\
   12 & 0.005354 &       32 &  0.030151 &          7 &  0.033964 &          6 &  0.035990 &          6 & 0.026365 &     12.75 \\
   17 & 0.012191 &       14 &  0.010846 &         19 &  0.010696 &         21 &  0.009280 &         25 & 0.010753 &     19.75 \\
   15 & 0.013121 &       12 &  0.010451 &         20 &  0.008462 &         24 &  0.005332 &         37 & 0.009342 &     23.25 \\
    2 & 0.000869 &       78 &  0.016206 &         13 &  0.017977 &         10 &  0.016756 &         11 & 0.012952 &     28.00 \\
    8 & 0.000688 &       99 &  0.020956 &         10 &  0.018782 &          9 &  0.004670 &         41 & 0.011274 &     39.75 \\
   11 & 0.002034 &       49 &  0.003632 &         49 &  0.003726 &         49 &  0.005071 &         40 & 0.003616 &     46.75 \\
    5 & 0.003089 &       42 &  0.004433 &         45 &  0.002178 &         66 &  0.004030 &         46 & 0.003432 &     49.75 \\
    3 & 0.000534 &      125 &  0.007971 &         30 &  0.003724 &         50 &  0.003557 &         51 & 0.003946 &     64.00 \\
    6 & 0.000389 &      158 &  0.008211 &         27 &  0.009874 &         22 &  0.000954 &        117 & 0.004857 &     81.00 \\
    7 & 0.000221 &      194 &  0.006495 &         35 &  0.005858 &         35 &  0.001861 &         75 & 0.003609 &     84.75 \\
   13 & 0.000752 &       87 &  0.001901 &         71 &  0.001986 &         70 &  0.000848 &        130 & 0.001372 &     89.50 \\
    0 & 0.000265 &      187 &  0.006695 &         34 &  0.007989 &         25 &  0.000762 &        142 & 0.003928 &     97.00 \\
    1 & 0.000398 &      155 &  0.005212 &         42 &  0.000989 &        115 &  0.000672 &        157 & 0.001817 &    117.25 \\
    9 & 0.000208 &      197 &  0.001170 &        102 &  0.000603 &        166 &  0.000691 &        154 & 0.000668 &    154.75 \\
\bottomrule
\end{tabular}}
\label{table:TDA_RF_Feature_Imp}
\end{table*}

\begin{table*}
\caption{Gradient Boosting TDA Feature Importance}
\resizebox{\textwidth}{!}{
\begin{tabular}{rrrrrrrrrrr}
\toprule
 name &       gb\_imp &  gb\_rank &     gb\_imp\_1 &  gb\_rank\_1 &     gb\_imp\_2 &  gb\_rank\_2 &     gb\_imp\_3 &  gb\_rank\_3 &  avg\_imp &  avg\_rank \\
\midrule
   16 & 2.615949e-01 &        1 & 3.365172e-01 &          1 & 1.788097e-01 &          2 & 3.744219e-01 &          1 & 0.287836 &      1.25 \\
   14 & 1.525905e-01 &        2 & 4.138910e-02 &          6 & 2.044894e-01 &          1 & 2.008486e-02 &          9 & 0.104638 &      4.50 \\
    4 & 3.407584e-02 &        6 & 5.093473e-02 &          5 & 1.718133e-02 &         13 & 2.277583e-02 &          8 & 0.031242 &      8.00 \\
   10 & 3.322310e-02 &        7 & 5.843127e-02 &          4 & 9.666783e-03 &         21 & 8.671493e-02 &          3 & 0.047009 &      8.75 \\
    9 & 1.811378e-03 &       38 & 8.254497e-03 &         13 & 9.991055e-04 &         65 & 8.593028e-03 &         15 & 0.004915 &     32.75 \\
   17 & 3.225699e-04 &       78 & 1.122728e-03 &         51 & 2.647563e-02 &          8 & 6.483156e-04 &         48 & 0.007142 &     46.25 \\
   12 & 3.263257e-03 &       29 & 7.942677e-03 &         14 & 1.845400e-05 &        134 & 2.624634e-03 &         28 & 0.003462 &     51.25 \\
    5 & 3.134408e-04 &       79 & 3.226526e-03 &         28 & 8.243095e-04 &         73 & 3.620715e-05 &         79 & 0.001100 &     64.75 \\
   15 & 1.838513e-04 &       85 & 1.174390e-07 &        132 & 2.269648e-02 &         10 & 2.304248e-04 &         63 & 0.005778 &     72.50 \\
    0 & 1.888470e-03 &       34 & 2.274994e-03 &         35 & 9.701916e-07 &        165 & 3.487678e-04 &         56 & 0.001128 &     72.50 \\
    2 & 1.882937e-04 &       84 & 1.087475e-03 &         52 & 2.042865e-03 &         47 & 3.227096e-08 &        137 & 0.000830 &     80.00 \\
    6 & 7.316263e-06 &      123 & 1.031175e-03 &         54 & 1.140362e-06 &        163 & 3.371906e-03 &         25 & 0.001103 &     91.25 \\
    8 & 2.664734e-05 &      111 & 3.140498e-06 &        105 & 1.040844e-03 &         64 & 6.738166e-06 &        102 & 0.000269 &     95.50 \\
    7 & 6.130102e-05 &       98 & 1.854843e-06 &        111 & 5.076095e-04 &         87 & 8.198926e-06 &         99 & 0.000145 &     98.75 \\
   11 & 5.479233e-05 &      100 & 5.192072e-04 &         67 & 1.720756e-04 &        105 & 1.443216e-09 &        145 & 0.000187 &    104.25 \\
    1 & 2.913908e-07 &      141 & 8.612808e-06 &         98 & 1.069186e-03 &         63 & 3.986463e-08 &        136 & 0.000270 &    109.50 \\
    3 & 7.092141e-06 &      124 & 1.616030e-12 &        171 & 1.502068e-06 &        158 & 1.280977e-03 &         37 & 0.000322 &    122.50 \\
   13 & 5.774614e-05 &       99 & 3.201980e-09 &        151 & 5.165003e-06 &        146 & 4.522834e-06 &        105 & 0.000017 &    125.25 \\
\bottomrule
\end{tabular}}
\label{table:TDA_GB_Feature_Imp}
\end{table*}

\clearpage

\end{document}